\documentclass[10pt,journal,comsoc]{IEEEtran}
\usepackage[T1]{fontenc}
\usepackage[utf8]{inputenc}
\usepackage{amsmath,amssymb}
\usepackage{graphicx}
\usepackage{epstopdf}
\usepackage{iftex}
\usepackage{amsmath,amssymb}   
\usepackage{graphicx}          
\usepackage{pgfplots}          
\pgfplotsset{compat=1.17}
\usepackage{algorithm}         
\usepackage{algpseudocode}     
\algnewcommand{\IfThen}[2]{\State \algorithmicif\ #1\ \algorithmicthen\ #2}
\ifXeTeX
\makeatletter
\let\originalincludegraphics\includegraphics
\renewcommand{\includegraphics}[2][]{%
	\begingroup
	\filename@parse{#2}\def\epsextension{eps}%
	\ifx\filename@ext\epsextension
	\edef\imagefilename{\filename@area\filename@base-eps-converted-to.pdf}%
	\else\def\imagefilename{#2}\fi
	\originalincludegraphics[#1]{\imagefilename}%
	\endgroup}
\makeatother
\fi
\usepackage{xcolor}
\usepackage{booktabs,array}
\usepackage{cite}
\usepackage{algorithm}
\usepackage{algpseudocode}
\usepackage{microtype}
\usepackage[colorlinks=true,linkcolor=red,citecolor=red,urlcolor=black,breaklinks=true]{hyperref}
\hypersetup{pdftitle={Agentic Wireless Optimization: Algorithm Design and Multi-Timescale Network Operation},pdfsubject={IEEE Wireless Communications two-column manuscript},pdfkeywords={Agentic optimization, large language models, algorithm design, wireless networks}}
\newcommand{\doi}[1]{doi: \href{https://doi.org/#1}{\nolinkurl{#1}}}
\newcolumntype{P}[1]{>{\raggedright\arraybackslash}p{#1}}

\begin{document}
	\bstctlcite{IEEEcontrol}

	\title{Agentic Wireless Optimization: Workflow Automation, Algorithm Design, and Multi-Timescale Network Operations}
	
	\author{Yuan~Guo, Zixiang~Ren, Ruihua~Luo, Yilong~Chen, Ya-Feng~Liu, Xianghao~Yu, Derrick~Wing~Kwan~Ng,~\IEEEmembership{Fellow,~IEEE,} and Jie~Xu,~\IEEEmembership{Fellow,~IEEE}%
		\thanks{Yuan Guo, Ruihua Luo, Yilong Chen, and Jie Xu are with the School of Science and Engineering (SSE), the Shenzhen Future Network of Intelligence Institute (FNii-Shenzhen), and the Guangdong Provincial Key Laboratory of Future Networks of Intelligence, The Chinese University of Hong Kong (Shenzhen), Guangdong 518172, China (e-mail: \mbox{guoyuan@cuhk.edu.cn}; \mbox{ruihualuo@link.cuhk.edu.cn}; \mbox{yilongchen@link.cuhk.edu.cn}; \mbox{xujie@cuhk.edu.cn}). (\textit{Corresponding author: Jie Xu}.)}%
		\thanks{Zixiang Ren is with the Department of Electrical and Computer Engineering, National University of Singapore, Singapore (e-mail: \mbox{zixiang\_ren@nus.edu.sg}).}%
		\thanks{Ya-Feng Liu is with the School of Mathematical Sciences, Beijing University of Posts and Telecommunications, Beijing, China (e-mail: \mbox{yafengliu@bupt.edu.cn}).}%
		\thanks{Xianghao Yu is with the Department of Electrical Engineering, City University of Hong Kong, Hong Kong (e-mail: \mbox{alex.yu@cityu.edu.hk}).}%
		\thanks{Derrick Wing Kwan Ng is with the School of Electrical Engineering and Telecommunications, University of New South Wales, Sydney, NSW 2052, Australia (e-mail: \mbox{w.k.ng@unsw.edu.au}).}%
	}
	
	\maketitle
\begin{abstract}
	Optimization is fundamental to wireless communications, from characterizing channel capacity and designing transmission schemes to allocating network resources. However, conventional optimization workflows require substantial domain expertise and manual effort, particularly when addressing high-dimensional or nonconvex problems, optimizing black-box objectives, or coordinating interdependent decisions across multiple time scales. Recent advances in agentic artificial intelligence (AI), particularly the development of agents based on large language models (LLMs), offer new opportunities to address these challenges by integrating reasoning, tool use, and numerical feedback. This article presents an overview of agentic AI for wireless optimization, focusing on workflow automation, algorithm design, and multi-timescale network operations. First, LLM-based agents can streamline optimization workflows by automating problem formulation, code generation, numerical solution, and result verification in a closed loop. Second, they facilitate algorithm design by combining and configuring existing numerical methods and refining learning-based optimizers, leveraging numerical feedback to improve solution quality and computational efficiency for challenging nonconvex and black-box problems. Third, they support multi-timescale network operations by decomposing coupled optimization problems into interdependent subtasks and coordinating their solutions across time scales to satisfy network-level objectives and service requirements. A multicast beamforming case study demonstrates performance gains from agent-developed solvers, particularly in terms of execution time, and a network energy-saving example further illustrates how agents can coordinate dynamic base station activation, user association, and beamforming across time scales. Finally, we discuss open challenges and research opportunities in agentic wireless optimization.
\end{abstract}

	
	\begin{figure*}[t]
		\centering
		\includegraphics[width=0.80\textwidth]{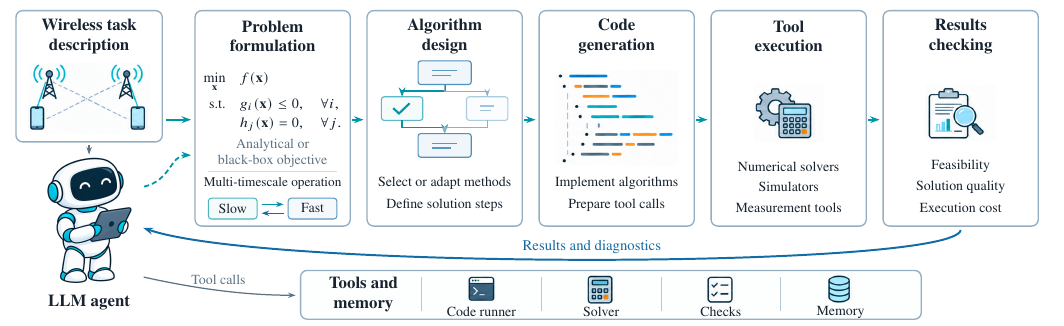}
		\caption{Closed-loop agentic workflow for wireless optimization. The LLM agent formulates the problem, designs algorithms, generates code, invokes tools, and checks results. Here, $\mathbf{x}$ denotes the decision variables, $f$ the objective function, and $g_i$ and $h_j$ the inequality and equality constraint functions, respectively. Objectives may be expressed analytically or evaluated as black boxes. For multi-timescale operation, slower configuration decisions govern faster allocation, while operational feedback informs subsequent configuration updates. Results and diagnostics guide revisions, while memory retains prior formulations, code, and outcomes.}
		\vspace{-2mm}
		\label{fig:workflow-automation}
	\end{figure*}

	\section{Introduction}
	\label{sec:introduction}
	
	Optimization plays a central role in wireless communications, from characterizing fundamental performance limits to designing and operating practical systems \cite{liu2024survey}. For example, characterizing channel capacity involves optimizing transmission strategies subject to resource constraints~\cite{tse2005fundamentals}, while wireless resource management requires optimizing beamforming, power control, user scheduling, and base station (BS) activation in response to time-varying channel conditions, traffic demands, and service requirements~\cite{ref27}.
	
	As wireless systems evolve toward sixth-generation (6G) networks, the associated optimization tasks become increasingly challenging. In practice, wider bandwidths, larger antenna arrays, denser deployments, and heterogeneous network elements increase problem dimensionality and introduce additional modeling complexity. The resulting problems are often highly nonconvex, with potentially competing objectives such as maximizing throughput, reducing energy consumption, and minimizing latency. Meanwhile, time-varying channels, user mobility, and fluctuating traffic demands, together with incomplete channel and network state information, further complicate accurate modeling and timely decision-making. In some scenarios, system performance can only be evaluated through simulations or measurements, giving rise to black-box optimization problems. Moreover, coupled network decisions must be meticulously coordinated across multiple time scales subject to stringent hardware constraints and computational deadlines. In fact, addressing these challenges via conventional optimization approaches requires substantial domain expertise and manual effort in properly formulating suitable optimization problems, selecting and implementing effective algorithms, and fine-tuning their parameters.

	Recent advances in agentic artificial intelligence (AI), particularly the development of agents based on large language models (LLMs), offer new opportunities to address these challenges through agentic wireless optimization. In this paradigm, an LLM-based agent combines LLM reasoning capabilities with external tools, such as code interpreters, numerical solvers, and simulators, while effectively leveraging memory of previous interactions~\cite{liu2025agenthpo}. Given a task description, the agent can formulate a mathematical optimization problem and identify suitable solution methods~\cite{ref3}. It then generates and executes the corresponding code and invokes tools to obtain numerical results. Numerical feedback guides iterative refinements of the problem formulation and solution algorithm~\cite{ref1}. Through this closed-loop process, the agent can systematically explore diverse formulations and algorithmic alternatives to identify effective solution methods with less manual effort~\cite{ref16}.

	Agentic wireless optimization draws on three complementary capabilities of LLM-based agents:
	\begin{itemize}
		\item \textbf{Workflow automation:} LLM-based agents can automate problem formulation, code generation, numerical solution, and result verification within a closed loop. A wireless optimization task expressed in natural language can thus be translated into a mathematical formulation, implemented as executable code, solved, and verified with reduced manual intervention.
		
		\item \textbf{Algorithm design:} The effectiveness of algorithms for nonconvex and black-box problems depends on the problem structure and operating conditions. Numerical feedback enables LLM-based agents to select, combine, and configure numerical methods to improve solution quality or reduce computational cost. Training and validation results similarly guide the agents in refining learning-based optimizers.
		
		\item \textbf{Multi-timescale network operations:} Wireless network operation involves coupled decisions across multiple time scales. LLM-based agents can decompose network-level optimization problems into coupled subtasks and coordinate their solutions, allowing decisions at slower time scales to account for the performance achievable through optimization at faster time scales.
	\end{itemize}
	
	Existing studies have explored the use of LLM-based agents to automate optimization workflows, including problem formulation, code generation, and solver execution~\cite{ref3,ref5,ref6}. Other studies have exploited their capabilities to generate candidate solutions and develop heuristic algorithms~\cite{ref1,ref16}. However, the systematic use of LLM-based agents to design and refine numerical solvers for challenging nonconvex and black-box wireless problems, as well as to coordinate coupled network decisions across multiple time scales, remains insufficiently explored. These gaps motivate a unified perspective on workflow automation, algorithm design, and multi-timescale network operations as three complementary aspects of agentic wireless optimization.

	To this end, this paper presents a comprehensive overview of agentic wireless optimization. First, we examine how LLM-based agents automate optimization workflows and reduce manual effort. Next, we explore agent-aided algorithm design, including solver orchestration for nonconvex problems, the adaptation of search methods for black-box objectives, and the training and refinement of learning-based optimizers. A multicast beamforming case study illustrates how numerical feedback guides the selection, combination, and configuration of numerical methods to develop efficient solvers. We then extend the discussion to multi-timescale network operations, using a conceptual network energy-saving example to illustrate how agents decompose coupled optimization tasks and coordinate their solutions. Finally, we discuss open challenges and research opportunities in trustworthy formulation and verification, wireless knowledge and tool integration, the cost and timeliness of agentic decisions, and benchmarking.
	\begin{figure*}[t]
		\centering
		\includegraphics[width=0.7\textwidth]{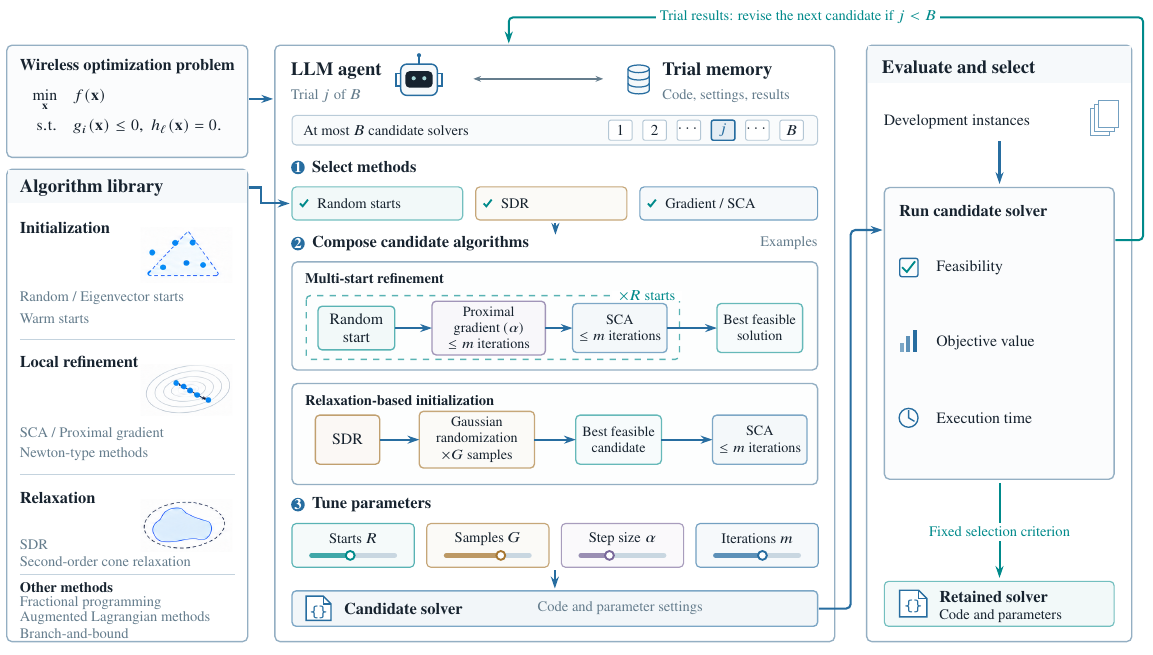}
		\caption{Agentic orchestration of solvers for wireless optimization problems. The agent evaluates at most $B$ candidate solvers, with $j$ denoting the current development trial. In the illustrative combinations, $R$ is the number of starting points, $G$ the number of Gaussian samples, $\alpha$ the proximal-gradient step size, and $m$ the iteration limit for each refinement method.}
		\vspace{-2mm}
		\label{fig:solver-orchestration}\label{fig:workflows}\label{fig:offline-agent-roles}
	\end{figure*}
	\section{Agentic Automation of Wireless Optimization Workflow}
	\label{sec:advances}
	
	LLM-based agents have attracted growing interest as an effective means of automating optimization workflows across domains such as operations research and machine learning, laying the foundation for their applications in wireless communications. Fig.~\ref{fig:workflow-automation} illustrates a closed-loop framework in which an LLM-based agent integrates problem formulation, algorithm design, code generation, tool execution, and result checking. Within this framework, LLM-assisted optimization can be broadly categorized into four complementary roles: direct solution generation, mathematical modeling and solver integration, algorithm development, and the training and refinement of learning-based optimizers~\cite{liu2025agenthpo,ref3,ref1,ref16}.
	
	First, LLMs can directly generate candidate solutions based on natural-language descriptions of optimization problem instances. For example, OPRO~\cite{ref1} uses previously evaluated candidate solutions and their objective values to guide an LLM in iteratively proposing new candidates. However, directly generated solutions may lack numerical precision or violate constraints, particularly for large-scale problems with tightly coupled decision variables.
	
	Second, LLM-based agents can translate natural-language descriptions into mathematical models and interface with numerical solvers~\cite{ref3}. For example, OptiMUS~\cite{ref3} formulates linear and mixed-integer linear programs from natural-language descriptions and automatically generates and debugs solver code. In particular, it reports a relative improvement of more than $30\%$ in modeling accuracy compared with prior methods on challenging benchmarks. The reliability of this approach, however, depends critically on the fidelity of the generated mathematical model. For instance, an omitted constraint or a misinterpreted parameter may lead to a numerically valid solution that does not satisfy the requirements of the intended optimization problem.
	
	Third, LLM-based agents can support the development of algorithms and solvers by generating or modifying code, evaluating its numerical performance, and leveraging numerical feedback to guide subsequent revisions. For example, ReEvo~\cite{ref16} combines evolutionary search with LLM-generated reflections to develop heuristics for NP-hard combinatorial optimization problems, achieving performance competitive with state-of-the-art heuristics on the evaluated benchmarks. In this case, the output is a reusable algorithm, making correctness verification and evaluation of generalization beyond the development instances essential.
	
	Fourth, LLM-based agents can support the development of learning-based optimizers, which employ neural networks to produce optimization solutions. Such networks can, for example, learn to mimic iterative algorithms, as demonstrated in wireless power allocation~\cite{sun2018learning}. In this setting, LLM-based agents can effectively coordinate training experiments and use their outcomes to refine subsequent configurations. For example, AgentHPO~\cite{liu2025agenthpo} leverages previous trials to guide hyperparameter search, achieving performance comparable to or better than that obtained through human-guided tuning on the evaluated machine-learning benchmarks. For wireless applications, maintaining constraint satisfaction and solution quality under changing network conditions remains essential.
	
	Motivated by these advances, various studies have adapted these capabilities to domain-specific optimization tasks in wireless communications. For example, LLM-OptiRA~\cite{ref5} constructs tractable reformulations or approximations for nonconvex resource-allocation problems and generates solver code with automated execution-error correction. Furthermore, ComAgent~\cite{ref6} coordinates specialized agents in a closed loop to translate high-level intents into solver-ready formulations and simulations, achieving performance comparable to that of expert-designed methods in its beamforming case study. In addition, WARA~\cite{ref22,guo2026trustworthy} automates the wireless optimization research process from problem proposal through modeling, algorithm design, and experimentation, adopting gate checks to assess consistency across stages and trigger targeted repairs.
	
	These studies demonstrate how LLM-based agents integrate problem formulation, implementation, and evaluation to automate wireless optimization workflows. Although such automation can reduce manual effort and improve workflow efficiency, these benefits alone do not necessarily establish superior solver performance over expert-designed algorithms. Therefore, an important research question is whether LLM-based agents, guided by numerical feedback, can systematically select, combine, and refine numerical methods to efficiently develop practical wireless solvers that improve solution quality or reduce computational cost relative to strong established baselines. This motivates the discussion of agent-aided algorithm design in the next section.
	
	\begin{figure*}[t]
		\centering
		\includegraphics[width=0.7\textwidth]{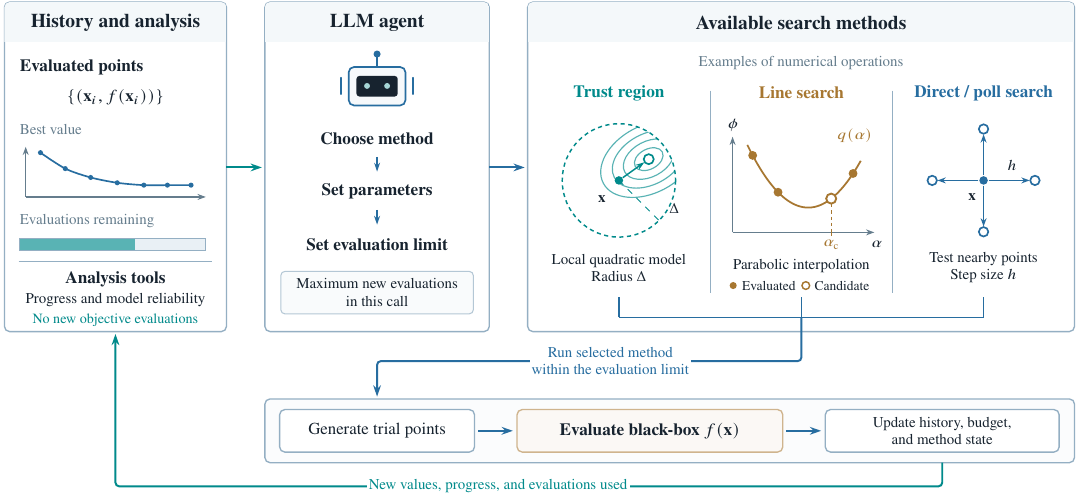}
		\caption{Agentic search for black-box optimization.}
		\vspace{-3mm}
		\label{fig:blackbox-agent}
	\end{figure*}

	\begin{figure*}[t]
		\centering
		\includegraphics[width=0.8\textwidth,trim={0pt 3pt 0pt 6pt},clip]{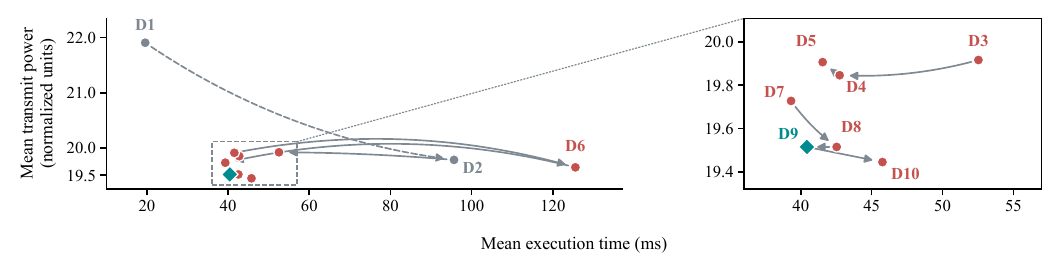}
		\caption{Development of the multicast beamforming solver. Each point represents a complete candidate solver and reports its mean transmit power and recorded execution time on the same $24$ development channels. D9 is retained and presented in Algorithm~\ref{alg:multicast}.  Arrows indicate evaluation order; the right panel enlarges the later candidates. D1, eigenvector initialization followed by SCA; D2, D1 with three additional smoothed random starts; D3, one restart and conditional SCA; D4, smoothed eigenvector initialization; D5, adjusted smoothing; D6, SDR followed by SCA; D7, selective SDR and restarts; D8, reordered steps and an additional restart; D9, selective restarts and SCA (retained); D10, extra random restarts. Recorded times include cached component costs and policy overhead, excluding LLM inference.}
		\label{fig:multicast}\label{fig:multicast-development-trajectory}
	\end{figure*}

	\section{Agentic Algorithm Design for Wireless Optimization}
	\label{sec:design}\label{sec:offline}
	
	This section examines how LLM-based agents facilitate the design of wireless optimization algorithms through three complementary roles: orchestrating numerical solvers for nonconvex problems, guiding derivative-free search for black-box objectives, and managing the training of learning-based optimizers. Across these roles, LLM-based agents adopt numerical feedback to select and combine optimization methods and tune their parameters, with the aim of improving solution quality and reducing computational cost. A multicast beamforming case study in Section~\ref{sec:multicast-case} then illustrates solver orchestration.
	
	\subsection{Agentic Orchestration of Solvers}
	\label{sec:composition}
	
	Wireless system design frequently involves nonconvex optimization problems in beamforming, power control, and resource allocation~\cite{liu2024survey}. Although globally optimal solutions are often difficult to obtain, various numerical methods can typically produce high-quality solutions. A representative class is nonconvex quadratically constrained quadratic programs (QCQPs), which commonly arise in transmit beamforming for multi-antenna wireless systems~\cite{sidiropoulos2006multicast}. Convex relaxation methods, particularly semidefinite relaxation (SDR), provide bounds on the optimal objective value and can be coupled with recovery procedures, such as Gaussian randomization, to generate feasible solutions. However, solving large SDR problems can incur substantial computational cost~\cite{sidiropoulos2006multicast}. Alternatively, local optimization methods, including successive convex approximation (SCA) and gradient-based algorithms~\cite{zhou2017coordinated}, can iteratively refine candidate solutions. However, their performance is highly dependent on initialization, and they generally lack global optimality guarantees. Furthermore, algorithmic parameters, such as the number of random starts, step sizes, and iteration limits, significantly affect the tradeoff between solution quality and computational cost.
	
	Effective solver design therefore involves selecting suitable numerical methods, determining their composition and execution order, and tuning their parameters to suit the problem structure and operating conditions. For example, candidate solutions obtained through convex relaxation and solution recovery can serve as initial points for subsequent refinement using local methods. In practice, the design space grows rapidly with the number of available methods and parameter choices, and assessing practical performance typically requires extensive numerical experiments. The results of these experiments then guide subsequent choices of method combinations and parameter settings. Manually implementing, comparing, and revising candidate designs is labor-intensive and can restrict exploration to a few familiar approaches.

	LLM-based agents can help automate this process by orchestrating numerical methods through an algorithm library, execution tools, and memory, as illustrated in Fig.~\ref{fig:solver-orchestration}. The library provides implementations of established methods, such as SDR and SCA, with configurable parameters. The agent constructs a candidate solver by selecting numerical methods, configuring their parameters, and specifying their execution order and invocation conditions. These conditions determine which numerical steps are executed for a given problem instance. Execution tools evaluate each candidate solver on a common set of problem instances and report feasibility, solution quality, and execution time. Memory records the tested solver designs and their outcomes, enabling the agent to compare alternatives and guide subsequent revisions. Through this iterative process, the agent can systematically explore the design space within a given evaluation budget, seeking improved solution quality and computational efficiency. The resulting reusable solver specifies the composition, execution order, parameter settings, and invocation conditions of established numerical methods. Once developed, it can be applied to previously unseen problem instances without requiring real-time LLM inference.

	\subsection{Agentic Search for Black-Box Optimization}
	\label{sec:search}
	
	Agentic optimization offers a promising approach to black-box optimization in wireless communications, where objectives generally lack accessible closed-form expressions and are evaluated through simulation or measurement without derivative information in practice. Since each evaluation consumes substantial resources, efficiently utilizing a limited evaluation budget is essential. Leveraging information from previous evaluations can help identify promising search directions and guide subsequent optimization steps. Direct-search methods, such as polling methods, compare objective values at trial points along selected directions, whereas model-based derivative-free optimization methods construct local surrogate models from previous evaluations to guide subsequent search.
	
	LLM agents can adaptively coordinate these approaches according to observed progress, as illustrated in Fig.~\ref{fig:blackbox-agent}. For example, NextEval~\cite{nexteval} connects an LLM agent with a numerical optimization environment providing analysis tools, model-based trust-region methods, direct-search polling steps, and geometry-repair operations for improving the local interpolation sample set. At each decision step, the agent selects an appropriate numerical operation, adjusts parameters such as step size, and allocates evaluations among methods within the remaining budget. The agent remains involved throughout the search for each problem instance and returns the best solution found within the evaluation budget. In contrast, the offline solver orchestration in Section~\ref{sec:composition} produces a reusable solver for subsequent problem instances.
	
	Numerical analysis supports these decisions before additional costly evaluations are performed. Using previously collected observations, analysis tools summarize optimization progress, assess local-model readiness and reliability, and predict objective changes associated with candidate steps. When a local model is sufficiently reliable, the agent can favor model-based steps. Otherwise, it can request additional samples to improve the model or switch to direct-search steps that do not rely on the model. The shared evaluation history and each method's numerical state are retained across tool calls, allowing accumulated information and numerical work to be reused when the corresponding methods are revisited.

	The experiments reported in~\cite{nexteval} consider $92$ unconstrained optimization problems with up to five variables, using perturbed starting points and rotated coordinates. Each run is limited to $50n$ objective evaluations, where $n$ denotes the problem dimension. Evaluation efficiency is assessed by the number of objective evaluations required to achieve specified relative objective-reduction targets. It is shown that NextEval achieves the highest aggregate evaluation-efficiency score among the compared methods, including model-based trust-region solvers and a quasi-Newton method with finite-difference gradients. These results highlight the potential of agent-guided search for wireless black-box optimization, where reaching a target performance level with fewer evaluations can reduce the required simulations or field measurements.

	\subsection{Agentic Training of Learning-Based Optimizers}
	\label{sec:learned}
	
	Learning-based optimizers provide an alternative approach to wireless optimization by training neural networks to map channel observations and system requirements directly to optimization solutions. Common approaches include imitating the outputs of iterative algorithms, as in power allocation~\cite{sun2018learning}, and unfolding the iterations of optimization algorithms into neural network layers with trainable parameters. Once trained, these optimizers can generate approximate solutions through a forward pass, reducing reliance on conventional iterative solvers during deployment. However, selecting effective architectures, hyperparameters, loss functions, and training distributions requires repeated training, testing, and validation. For example, minimizing deviations from reference solutions does not necessarily optimize the underlying communication objective, while adjusting the training distribution may improve performance under some channel conditions but degrade it under others. These interacting choices make optimizer development computationally costly.

	LLM-based agents can facilitate systematic experimentation by using training diagnostics and performance metrics to guide subsequent runs. Besides aggregate validation scores, these agents can examine achieved communication rates and constraint violations across channel conditions to identify performance limitations. These observations help them formulate hypotheses about performance gaps and select targeted changes to test. While numerical optimizers update neural network weights within each run, LLM-based agents adapt training configurations across runs based on previous results. AgentHPO~\cite{liu2025agenthpo} illustrates this experimental loop specifically through agent-guided hyperparameter search. Extending this approach to wireless optimization remains an emerging research direction. Key open questions include efficiently allocating limited training resources among candidate configurations and ensuring constraint satisfaction and consistent solution quality when channel statistics and network conditions change after deployment.

	\begin{figure*}[t]
		\centering
		\includegraphics[width=0.9\textwidth]{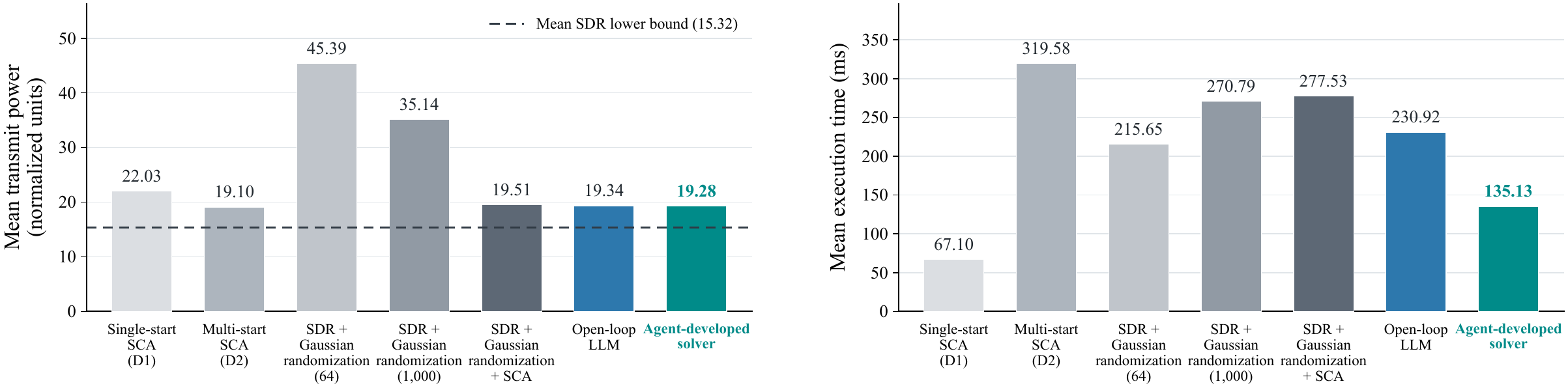}
		\caption{Mean transmit power (left) and mean execution time (right) on $96$ held-out channels; the dashed line denotes the mean certified SDR lower bound. The channels evenly cover $12$ configurations with $N\in\{8,16\}$ antennas, $K\in\{12,24,48\}$ users, and $\rho\in\{0,0.8\}$, with unit noise power and a $10$-dB SNR target. Specifically, $\mathbf h_k\sim\mathcal{CN}(\mathbf0,\ell_k\mathbf R)$, where $R_{mn}=\rho^{|m-n|}$, $\ell_k=10^{-a_k/10}$, and independent $a_k\sim\mathcal{U}[0,6]$~dB; channel vectors are not normalized to unit norm. D1 applies up to $8$ SCA iterations from the principal-eigenvector initialization; D2 adds $3$ smoothed random starts with up to $16$ SCA iterations each. For SDR methods, parenthesized numbers count Gaussian candidates, and a principal-eigenvector candidate is also included. SDR--SCA uses $64$ Gaussian candidates and up to $8$ SCA iterations on the best feasible beamformer. Timing averages $3$ interleaved single-thread repetitions of complete solver calls for all methods in the same batch on a common host, excluding LLM inference.}
		\label{fig:offline-case-study}
	\end{figure*}
	
	\begin{figure*}[t]
		\centering
		\includegraphics[width=0.7\textwidth]{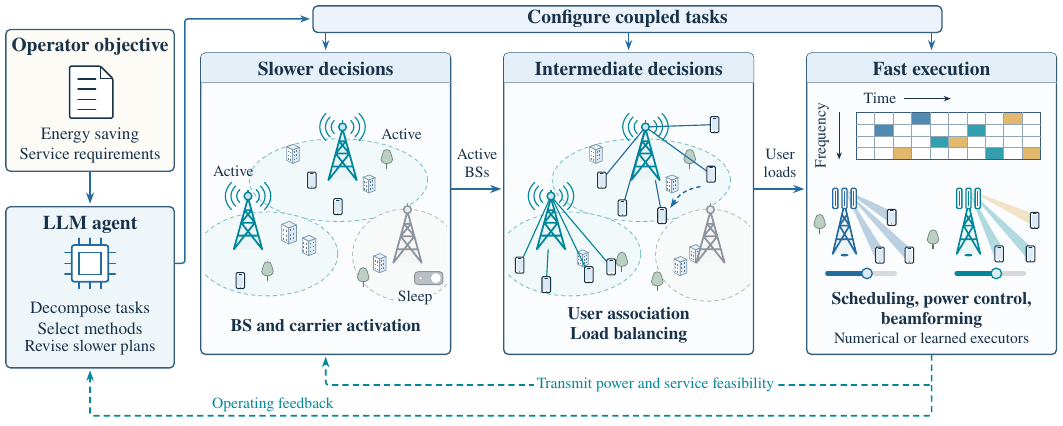}
		\caption{Agentic coordination for network energy saving across time scales. Activation and association decisions are passed to faster tasks, while the resulting transmit power and service feasibility are fed back to revise them.}
		\vspace{-3mm}
		\label{fig:online-adaptation}
	\end{figure*}

	\subsection{Case Study of Multicast Beamforming}
	\label{sec:multicast-case}
	
	To illustrate the potential benefits of agentic solver orchestration, we consider a single-group multicast beamforming problem, in which a base station minimizes its transmit power while satisfying every user's signal-to-noise ratio (SNR) requirement. This problem is generally NP-hard and is a classical example where relaxation and local optimization methods are complementary~\cite{sidiropoulos2006multicast}. Accordingly, the numerical library includes SDR for obtaining a transmit-power lower bound, Gaussian randomization with feasibility scaling to construct feasible candidate beamformers, and SCA for local refinement. Initialization and smoothing routines~\cite{zhou2017coordinated} are employed to generate and improve initial beamforming directions for SCA.
	
	To isolate the effect of numerical feedback, the agent and an open-loop LLM use the same model, numerical library, and budget of $10$ candidate solvers, including $2$ common initial candidates.\footnote{Both use \texttt{gpt-6.1-sol} with \texttt{ultra} reasoning.} Both can select methods, configure parameters and execution order, and define invocation conditions using the antenna count $N$, user count $K$, channel spectral concentration $e$, and intermediate solution statistics. The open-loop LLM specifies all remaining candidates before receiving numerical results, whereas the agent uses feedback to revise subsequent candidates. Each candidate, including its conditions and thresholds, is fixed before evaluation on the same $24$ development channels. Among candidates satisfying all SNR constraints on these channels, each controller retains the solver minimizing a common score $J$ that combines normalized transmit power and execution time.\footnote{The score is $J=\frac{1}{M}\sum_{i=1}^{M}(P_i/L_i+t_i)$, where $i$ indexes the $M=24$ development channels, $P_i$ is the transmit power, $L_i$ is the certified SDR lower bound, and $t_i$ is the recorded execution time in seconds. The SDR lower bounds are computed from dual solutions after rescaling the nonnegative dual multipliers and verifying dual feasibility within a numerical tolerance of $10^{-9}$.}

	\begin{algorithm}[t]
		\caption{Retained Multicast Beamforming Solvers}
		\label{alg:multicast}
		\footnotesize
		\begin{algorithmic}[1]
			\Require Channels $\mathbf H=[\mathbf h_1,\ldots,\mathbf h_K]\in\mathbb C^{N\times K}$, SNR target $\gamma$, noise power $\sigma^2$
			\Statex \textbf{Notation:} $e=\lambda_{\max}(\mathbf H^H\mathbf H)/\mathrm{tr}(\mathbf H^H\mathbf H)$ measures the spectral concentration of the channel matrix; $r=\|\mathbf w^\star\|^2/\max_k(\gamma\sigma^2/\|\mathbf h_k\|^2)$ is the current power relative to the single-user bound. Each ``try'' keeps only feasible, lower-power candidates. Update $r$ whenever $\mathbf w^\star$ changes.
			\Statex Execute the two solvers independently.
			\Statex \textbf{Agent-developed solver (D9)}
			\State $\mathbf w^\star\gets$ SCA from the smoothed principal-eigenvector initialization
			\IfThen{$N\geq16$ and $e\geq0.16$}{try SCA from a smoothed random initialization}
			\IfThen{$N\leq8$ and $K\geq48$}{try SCA from a smoothed random initialization}
			\IfThen{$N\leq8$, $K\leq12$, and $e\leq0.45$}{try SCA from the SDR solution with Gaussian randomization}
			\IfThen{$N\leq8$, $K\geq48$, and $e\leq0.30$}{try SCA from another smoothed random initialization}
			\IfThen{$N\leq8$, $K=24$, and $e\leq0.45$}{try additional SCA iterations from~$\mathbf w^\star$}
			\Statex \textbf{Open-loop LLM solver (O)}
			\setcounter{ALG@line}{0}
			\State $\mathbf w^\star\gets$ SCA from the principal-eigenvector initialization
			\IfThen{$e\geq0.6$ and $r\geq1.5$}{try SCA from a random initialization}
			\IfThen{$e<0.6$ and $r\geq1.8$}{try SCA from a smoothed random initialization}
			\IfThen{$K\geq24$ and $r\geq2$}{try SCA from a smoothed random initialization}
			\IfThen{$K\geq48$ and $r\geq2.5$}{try SCA from a smoothed random initialization}
			\IfThen{the last executed try reduces the power by more than $1\%$}{try additional SCA iterations from~$\mathbf w^\star$}
			\Ensure Beamformer $\mathbf w^\star$
		\end{algorithmic}
	\end{algorithm}

	Fig.~\ref{fig:multicast-development-trajectory} reports the mean transmit power and execution time of each complete candidate solver. To guide revisions, the agent also receives per-configuration results and the power reduction and execution cost of individual steps. Starting from D1, which applies SCA from the principal-eigenvector initialization, and D2, which adds smoothed random starts, the agent adjusts initialization and refinement in D3--D5. Using SDR-based initialization on all development channels in D6 reduces power but substantially increases execution time, with benefits varying across configurations. The agent therefore introduces conditional SDR execution in D7: after smoothing and SCA, SDR is attempted only when $N\leq8$, $K\leq12$, and $e\leq0.45$. Subsequent trials refine these execution conditions. An additional smoothed random start in D8 improves some eight-antenna configurations but not the corresponding sixteen-antenna configurations. The agent therefore adds $N\leq8$ to this step's invocation conditions in D9 and tests additional SCA refinement in other configurations. After all $10$ trials, $J$ selects D9 over D10, whose slightly lower power is outweighed by its longer execution time. Algorithm~\ref{alg:multicast} presents the retained D9 and open-loop solver.
	
	Finally, the retained solvers are fixed and compared with D1, D2, and three SDR-based baselines on $96$ held-out channels within the same configurations. As shown in Fig.~\ref{fig:offline-case-study}, the agent-developed solver achieves lower mean transmit power and shorter execution time than SDR with Gaussian randomization alone. Compared with SDR--SCA and the open-loop solver, it maintains comparable power with substantially shorter execution time. D1 is faster but requires higher power, whereas D2 achieves slightly lower power at substantially greater execution cost. These comparisons show how numerical feedback guides selective computation to approach the solution quality of more expensive solvers.

	\section{Agentic Coordination for Multi-Timescale Network Operation}
	\label{sec:operation}
	
	Beyond individual algorithm design in Section~\ref{sec:design}, wireless network operation requires coordinating decisions across multiple time scales to achieve network-level objectives, such as minimizing energy consumption while satisfying service quality. At longer time scales, such as minutes to hours, the network adaptively controls the activation of BSs and carriers to track traffic variations, during which components within active BSs, such as power amplifiers, can enter sleep states during the idle periods. At intermediate time scales, on the order of seconds to minutes, the network optimizes user association and load balancing to determine which users are served by each active BS. At still shorter time scales, the active BSs design the user scheduling, power control, and beamforming in accordance with instantaneous channel and traffic conditions. These decisions are tightly coupled. For example, deactivating a BS during periods of light traffic reduces its operating power but requires reassociating its users with other BSs, which may need higher transmit power or fail to fulfill service requirements. Therefore, the energy benefit of an activation decision depends on the association and beamforming decisions it induces~\cite{ref27}.
	
	In practice, this coupling renders the coordination challenging. Specifically, the joint optimization combines discrete BS activation and user association with continuous resource allocation, yielding mixed-integer nonconvex problems that are difficult to tackle for medium- to large-scale networks~\cite{ref27}. Moreover, slower-time-scale decisions remain in effect as channels and traffic evolve, so evaluating them often requires solving faster subproblems under representative operating conditions. The formulation and decomposition also depend on the network objective, deployment, and available control functions. As such, changes in energy targets or service requirements may require reassessing the existing procedure and revising affected components. Developing and updating these procedures typically requires substantial domain expertise and manual effort.
	
	LLM-based agents offer a promising approach to automate the construction and revision of such coordinated procedures. Given an operator-specified objective, an LLM agent can use network information and BS power models to formulate a joint optimization problem over BS activation, user association, and beamforming. It then decomposes the problem by time scale, as illustrated in Fig.~\ref{fig:online-adaptation}. The slower layers pass the active BS configuration and user assignments to the beamforming layer, which reports the required transmit power and whether service requirements are satisfied. For each subproblem, the agent selects a numerical solver or learning-based optimizer, such as those discussed in Section~\ref{sec:design}. To evaluate a candidate activation pattern, it reassociates users and invokes the beamforming executor under representative operating conditions. Patterns for which no solutions satisfying the service requirements are obtained under these conditions are discarded. For the remaining patterns, network power consumption is estimated using the resulting transmit powers and BS power models. The activation decision thus accounts for both the power saved by sleeping BSs and the additional power consumed by those remaining active.
	
	After deployment, dedicated numerical solvers or learning-based optimizers, developed or configured with the assistance of LLM-based agents as discussed in Section~\ref{sec:design}, should handle time-critical decisions, such as scheduling and beamforming. The LLM-based agent then operates on the slower planning and revision loop, which could be implemented in non-real-time. Persistent variations in traffic or service requirements prompt the agent to reconsider activation and association, while recurrent deadline violations trigger a review of the relevant executor and its computational workload. Building on established methods for joint BS activation and beamforming~\cite{ref27}, the agent's role is to construct, coordinate, and revise the overall procedure according to the operator's objective.
	
	\section{Open Challenges and Research Opportunities}
	\label{sec:challenges}
	
	Despite recent progress, further research is needed to make agentic wireless optimization reliable, efficient, and rigorously evaluated in practice.
	
	\subsection{Trustworthy Formulation and Verification}
	
	Ensuring correctness is essential for deploying agent-generated algorithms and network configurations. LLM-based agents may produce seemingly plausible but mathematically incorrect formulations or code by omitting a power constraint, mixing linear and decibel units, or misinterpreting a service requirement. Such errors can remain undetectable because a solver may return a feasible solution to a mis-specified problem. Verification must therefore address both the fidelity to the intended requirements and numerical validity. In this regard, wireless structure provides useful checks, including physical consistency tests, simulation-based constraint checks, and bounds from duality or convex relaxation. An important research direction is to incorporate these mechanisms throughout the agent workflow, attach traceable evidence to generated results, and repeat relevant checks after revisions. For operating networks, proposed configurations should undergo validation before deployment and support rollback when observed performance fails to satisfy requirements.
	
	\subsection{Wireless Knowledge and Tool Integration}
	
	Agent performance also depends on access to wireless domain knowledge and dependable tools. Channel models, interference structures, and protocol constraints inform problem formulation and algorithm selection, while solvers, system-level simulators, and ray-tracing engines require interfaces with clearly defined inputs, outputs, units, and assumptions. Research opportunities include developing agent-accessible libraries, integrating relevant standards and literature, and retaining experience from previous optimization tasks. These records should document the formulations, method choices, parameter settings, outcomes, and operating conditions of each trial. Reusing these records can guide subsequent decisions, provided that their applicability is assessed and the resulting designs are validated under the current network conditions.
	
	\subsection{Cost and Timeliness of Agentic Decisions}
	
	The benefits of agentic optimization must be assessed against its end-to-end costs. LLM inference consumes time, computation, and energy, while tool execution, communication, and reconfiguration introduce additional overhead. For network operations, the agent's reasoning and tool interactions must align with the timing requirements of the decisions being revised. Time-critical decisions can be delegated to numerical solvers or learning-based optimizers, with agents handling slower planning and revision. Research is needed to determine when agent intervention is worthwhile, how to allocate tasks between large cloud-based models and lightweight edge models, and how frequently agents should intervene. Distributed deployment further requires coordination mechanisms that account for communication delays, outdated network state information, and potentially conflicting actions across cells and network layers.
	
	\subsection{Benchmarks for Agentic Wireless Optimization}
	
	Rigorous benchmarks are essential for measuring progress and identifying performance gains from agentic capabilities. Evaluations should report solution quality and constraint satisfaction alongside LLM inference costs, numerical trials, execution time, and human effort. Comparisons should clearly specify the resources matched across methods and account for any additional costs separately. Open-loop LLM-based approaches using the same model, tools, and candidate budget can help assess the value of numerical feedback, while strong expert-designed algorithms and classical numerical methods establish practical competitiveness. NextEval Bench~\cite{nexteval} provides an example for derivative-free optimization benchmarks. Developing shared tasks, tools, and evaluation protocols for representative wireless problems would support reproducible comparisons across agents. Within this framework, controlled ablation studies should assess the contributions of agent-level feedback, memory, and revision strategies while retaining the same numerical tools. These studies would clarify where agentic gains arise and guide the development of more effective and efficient agents.
	
	\section{Conclusions}
	\label{sec:conclusions}
	
	This paper presented an overview of agentic wireless optimization, highlighting how LLM-based agents combine reasoning, tool use, memory, and numerical feedback to support the formulation and solution of wireless optimization problems. We examined how these agents automate optimization workflows and support algorithm design through solver orchestration, black-box search, and the refinement of learning-based optimizers. We further considered their role in coordinating coupled network decisions across time scales and discussed open challenges and research opportunities. We envision broader applications of agentic optimization in wireless system design and network operation, with LLM-based agents working alongside wireless experts to develop and coordinate effective optimization methods with less manual effort.
	
	\bibliographystyle{IEEEtran}
	
	\bibliography{references}

\begin{thebibliography}{10}
\providecommand{\url}[1]{#1}
\csname url@samestyle\endcsname
\providecommand{\newblock}{\relax}
\providecommand{\bibinfo}[2]{#2}
\providecommand{\BIBentrySTDinterwordspacing}{\spaceskip=0pt\relax}
\providecommand{\BIBentryALTinterwordstretchfactor}{4}
\providecommand{\BIBentryALTinterwordspacing}{\spaceskip=\fontdimen2\font plus
\BIBentryALTinterwordstretchfactor\fontdimen3\font minus
  \fontdimen4\font\relax}
\providecommand{\BIBforeignlanguage}[2]{{%
\expandafter\ifx\csname l@#1\endcsname\relax
\typeout{** WARNING: IEEEtran.bst: No hyphenation pattern has been}%
\typeout{** loaded for the language `#1'. Using the pattern for}%
\typeout{** the default language instead.}%
\else
\language=\csname l@#1\endcsname
\fi
#2}}
\providecommand{\BIBdecl}{\relax}
\BIBdecl

\bibitem{liu2024survey}
Y.-F. Liu, T.-H. Chang, M.~Hong, Z.~Wu, A.~M.-C. So, E.~A. Jorswieck, and
  W.~Yu, ``A survey of recent advances in optimization methods for wireless
  communications,'' \emph{IEEE J. Sel. Areas Commun.}, vol.~42, no.~11, pp.
  2992--3031, Nov. 2024.

\bibitem{tse2005fundamentals}
D.~Tse and P.~Viswanath, \emph{Fundamentals of Wireless Communication}.\hskip
  1em plus 0.5em minus 0.4em\relax Cambridge, U.K.: Cambridge Univ. Press,
  2005.

\bibitem{ref27}
J.~Lin, M.~Ma, and X.~Wang, ``Joint design of long-term base station activation
  and short-term beamforming for green wireless networks,'' \emph{IEEE Trans.
  Wireless Commun.}, vol.~23, no.~1, pp. 185--200, Jan. 2024.

\bibitem{liu2025agenthpo}
S.~Liu, C.~Gao, and Y.~Li, ``{AgentHPO}: Large language model agent for
  hyper-parameter optimization,'' in \emph{Proc. Conf. Parsimony Learn.
  (CPAL)}, ser. Proc. Mach. Learn. Res., vol. 280, 2025, pp. 1146--1169.

\bibitem{ref3}
A.~Ahmaditeshnizi, W.~Gao, and M.~Udell, ``{OptiMUS}: Scalable optimization
  modeling with ({MI}){LP} solvers and large language models,'' in \emph{Proc.
  41st Int. Conf. Mach. Learn. ({ICML})}, ser. Proc. Mach. Learn. Res., vol.
  235.\hskip 1em plus 0.5em minus 0.4em\relax PMLR, 2024, pp. 577--596.

\bibitem{ref1}
C.~Yang, X.~Wang, Y.~Lu, H.~Liu, Q.~V. Le, D.~Zhou, and X.~Chen, ``Large
  language models as optimizers,'' in \emph{Proc. Int. Conf. Learn. Represent.
  ({ICLR})}, 2024.

\bibitem{ref16}
H.~Ye, J.~Wang, Z.~Cao, F.~Berto, C.~Hua, H.~Kim, J.~Park, and G.~Song,
  ``{ReEvo}: Large language models as hyper-heuristics with reflective
  evolution,'' in \emph{Proc. Adv. Neural Inf. Process. Syst. (NeurIPS)},
  vol.~37, 2024, pp. 43\,571--43\,608.

\bibitem{ref5}
X.~Peng, Y.~Lu, Y.~Cang, C.~Cao, and M.~Chen, ``{LLM-OptiRA}: {LLM}-driven
  optimization of resource allocation for non-convex problems in wireless
  communications,'' in \emph{Proc. {IEEE} Globecom Workshops ({GC Wkshps})},
  2025, pp. 1131--1136.

\bibitem{ref6}
H.~Li, M.~Xiao, K.~Wang, R.~Schober, D.~I. Kim, and Y.~L. Guan, ``{ComAgent}:
  Multi-{LLM} based agentic {AI} empowered intelligent wireless networks,''
  \emph{{IEEE} Wireless Commun.}, pp. 1--12, Jul. 2026, early access.

\bibitem{sun2018learning}
H.~Sun, X.~Chen, Q.~Shi, M.~Hong, X.~Fu, and N.~D. Sidiropoulos, ``Learning to
  optimize: Training deep neural networks for interference management,''
  \emph{IEEE Trans. Signal Process.}, vol.~66, no.~20, pp. 5438--5453, Oct.
  2018.

\bibitem{ref22}
Y.~Guo, Y.~Chen, C.~Hu, X.~Yu, L.~Hong, and J.~Xu, ``{WARA}: Toward automated
  wireless optimization research with closed-loop {LLM} agents,'' \emph{IEEE
  Open J. Commun. Soc.}, vol.~7, pp. 13\,509--13\,524, Oct. 2026.

\bibitem{guo2026trustworthy}
Y.~Guo, Z.~Ren, J.~Xu, L.~Hong, F.~Liu, and R.~Zhang, ``Exploiting {LLM} agents
  for trustworthy {AutoResearch} in wireless communications,'' arXiv preprint
  arXiv:2609.06174, 2026.

\bibitem{sidiropoulos2006multicast}
N.~D. Sidiropoulos, T.~N. Davidson, and Z.-Q. Luo, ``Transmit beamforming for
  physical-layer multicasting,'' \emph{IEEE Trans. Signal Process.}, vol.~54,
  no.~6, pp. 2239--2251, Jun. 2006.

\bibitem{zhou2017coordinated}
L.~Zhou, L.~Zheng, X.~Wang, W.~Jiang, and W.~Luo, ``Coordinated multicell
  multicast beamforming based on manifold optimization,'' \emph{IEEE Commun.
  Lett.}, vol.~21, no.~7, pp. 1673--1676, Jul. 2017.

\bibitem{nexteval}
S.~Chen, L.~Cui, C.~Huang, Y.-F. Liu, M.~Xie, M.~Xu, and Z.~Zhang,
  ``{NextEval}: An agent-based solver and benchmark for derivative-free
  optimization,'' arXiv preprint arXiv:2610.36553, Oct. 2026.

\end{thebibliography}
	
\end{document}